\documentclass{Interspeech}

\interspeechcameraready

\title{Enhancing Retrieval-Augmented Audio Captioning with Generation-Assisted Multimodal Querying and Progressive Learning}

\author[affiliation={1}]{Changin}{Choi}
\author[affiliation={1}]{Sungjun}{Lim}
\author[affiliation={1,2}]{Wonjong}{Rhee}

\affiliation{IPAI}{Interdisciplinary Program in Artificial Intelligence, Seoul National University}{Korea}
\affiliation{}{Department of Intelligence and Information, Seoul National University}{Korea}
\email{ci2015.choi@snu.ac.kr, lsjung567@snu.ac.kr, wrhee@snu.ac.kr}

\usepackage{comment}
\usepackage{newfloat}
\usepackage{listings}
\usepackage[table]{xcolor}
\usepackage{colortbl}
\usepackage{makecell}
\usepackage{booktabs}
\usepackage{multirow}
\usepackage{amsmath}
\usepackage{amssymb}
\usepackage{graphicx}
\usepackage{caption}
\usepackage{etoolbox}
\usepackage{float}

\begin{document}

\maketitle

\begin{abstract}
Retrieval-augmented generation can improve audio captioning by incorporating relevant audio-text pairs from a knowledge base.  Existing methods typically rely solely on the input audio as a unimodal retrieval query. In contrast, we propose Generation-Assisted Multimodal Querying, which generates a text description of the input audio to enable multimodal querying. This approach aligns the query modality with the audio-text structure of the knowledge base, leading to more effective retrieval. Furthermore, we introduce a novel progressive learning strategy that gradually increases the number of interleaved audio-text pairs to enhance the training process.  Our experiments on AudioCaps, Clotho, and Auto-ACD demonstrate that our approach achieves state-of-the-art results across these benchmarks.
\end{abstract}

\section{Introduction}
\label{sec:introduction}

Recent advancements in audio understanding tasks have been driven by large language models (LLMs), where increasing model size and training on larger datasets have led to significant improvements. However, continually scaling model size and training data poses substantial computational burdens. 
As an alternative, retrieval-augmented audio captioning has gained traction. In this approach, audio query is used to search for relevant audio-text pairs using either audio-to-audio~\cite{kong2024audioflamingo} or audio-to-text~\cite{2024recap} similarity. The retrieved audio-text pairs are provided with the audio query, enabling the audio language model to generate more accurate captions. 

\begin{figure}[t]
    \centering
    \includegraphics[width=\linewidth]{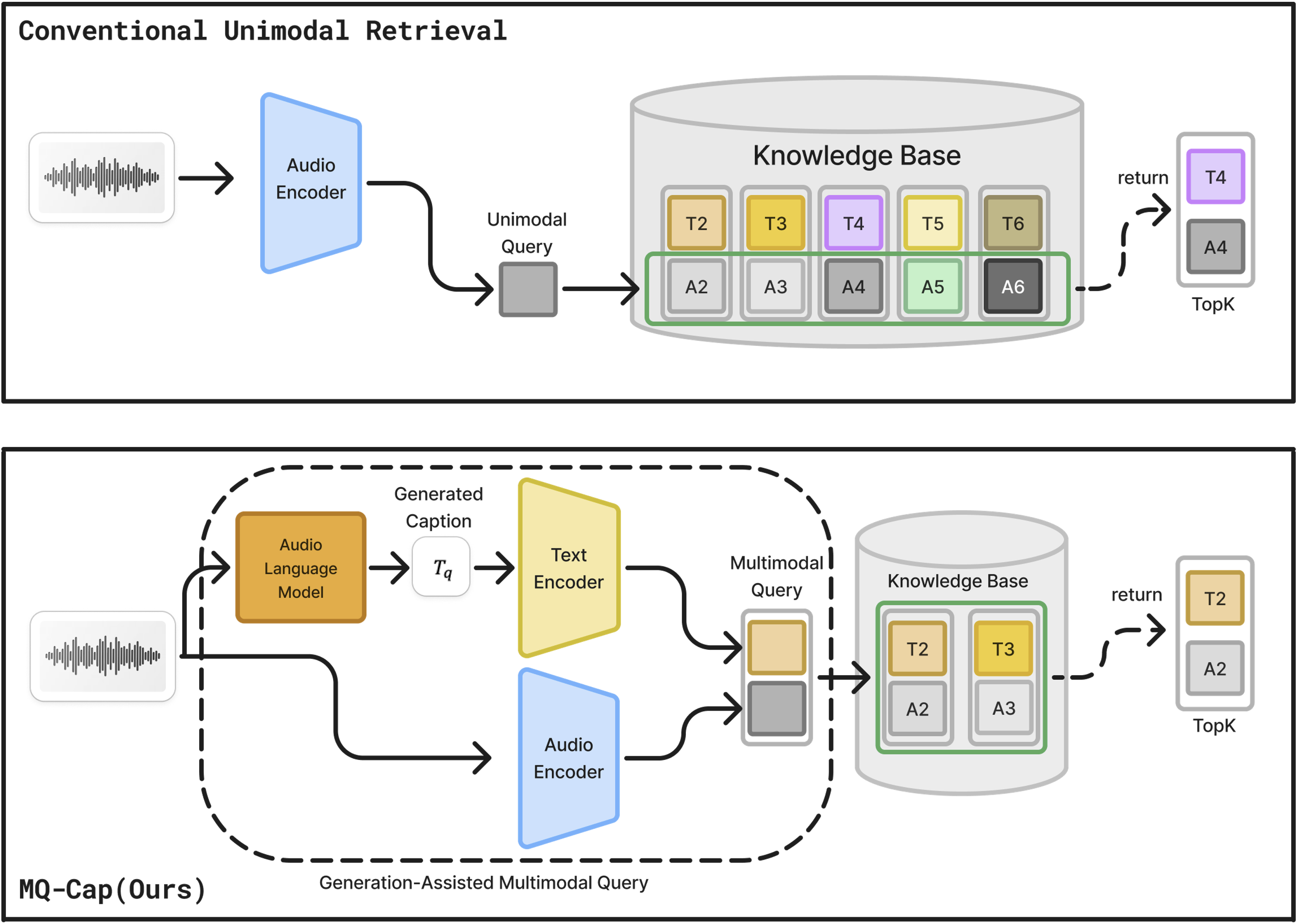}
    \caption{Illustration of how MQ-Cap differs from conventional approaches. MQ-Cap leverages \textit{Generation-Assisted Multimodal Query}~(GAMQ), where the audio-language model generates a caption \( T_q \) from the audio query \( A_q \), effectively transforming the query into a multimodal representation by incorporating both the original audio and the generated caption. 
    }
    \label{fig:pair_to_pair_retrieval}
\end{figure}
\setlength{\floatsep}{5pt} 
\setlength{\textfloatsep}{5pt} 

To generate accurate responses, retrieval-augmented generation relies on the relevance of retrieved examples. However, retrieval models often return irrelevant or noisy data, which can distract the language model from generating precise responses~\cite{petroni2020how, Wu2024HowED}. To mitigate these challenges, recent advances in query transformation consider refining or expanding queries to enhance retrieval quality \cite{ma2023query, gao2022precise}. Their application to audio captioning, however, remains under-explored. In this work, we develop a tailored query transformation method for retrieval-augmented audio captioning.

In our approach, we introduce a novel query transformation that converts a unimodal audio query into a multimodal audio-text query. Unimodal querying is inherently limited, as it computes similarity within or across modalities without fully capturing their relationships, often resulting in misalignment issues~\cite{kang2023noise, luong2024revisiting}. Traditionally, this limitation has been overlooked because only an audio query is provided, without its corresponding text counterpart. To address this, we generate a caption from the audio query, creating an audio-text pair for retrieval. We will show through experiments that multimodal querying with a generated caption can significantly improve RAG-based audio captioning.

Since retrieval-augmented audio captioning utilizes interleaved sequences of the retrieved pairs, it is essential to improve few-shot performance. We employ progressive learning through interleaved audio–text sequences to effectively leverage retrieved knowledge~\cite{yasunaga2022retrieval}. We employ a two-stage training process in which Stage 2 progressively increases the count of interleaved audio–text pairs to improve few-shot performance.

We propose MQ-Cap (\textbf{M}ultimodal \textbf{Q}uerying based \textbf{Cap}tioning), illustrated in Figure~\ref{fig:pair_to_pair_retrieval}, which offers the following contributions. 
\begin{itemize} 
\item We propose Generation-Assisted Multimodal Querying~(GAMQ) approach that leverages generated captions to construct multimodal queries, enhancing the retrieval.
\item We adopt a progressive learning approach to improve the few-shot training of interleaved audio-text pairs.
\end{itemize}

\section{Method}
\label{sec:method}
%
In this section, we first describe the architectural components and training process of the base model, which incorporates progressive learning. Next, we explain the pairwise retrieval process using a multimodal query, including a simple formulation for measuring multimodal similarity. 


\subsection{Base model for RAG}

\begin{figure}[t]
    \centering
    \includegraphics[width=0.9\linewidth]{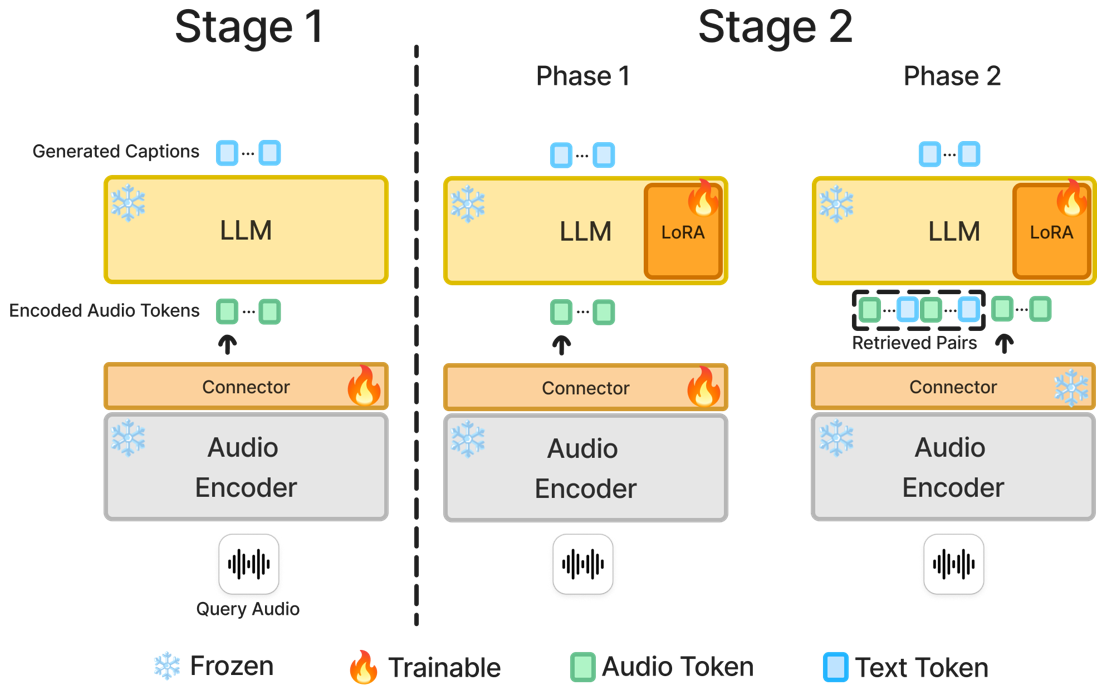}
    \caption{Illustration of the architectural components and training process of the base model. During stage 1, the MLP connector is optimized to align audio features with the language model. In stage 2, the number of retrieved pairs is progressively increased to enhance the model’s ability to generate captions in few-shot settings.
    }
    \label{fig:model_architecture}
\end{figure}


%
The architecture of our base model and its training process are illustrated in Figure \ref{fig:model_architecture}. To encode query audio, we use Laion-CLAP~\cite{2023laionclap}, a widely used audio encoder known for generating robust and meaningful audio representations. These audio features are subsequently transformed into audio tokens using a multi-layer perceptron (MLP) connector. The audio tokens are then processed by Vicuna~\cite{vicuna2023}, a widely used instruction-tuned language model, to generate corresponding text descriptions.

The training process consists of two stages. In stage 1, the MLP connector is trained to align the audio features with the language model. In stage 2, we fine-tune the MLP connector and LoRA adapters~\cite{hu2022lora} of Vicuna to handle interleaved audio–text pairs in a few-shot setting. In stage 2, we adopt a progressive learning approach, dividing training into two phases to gradually enhance the model’s ability to handle interleaved audio-text pairs. In the first phase, the model is trained to generate captions for query audio without interleaved pairs. In the second phase, model is trained using interleaved audio-text sequences with a progressively increasing number of pairs to enhance the model's few-shot captioning capabilities. During training, we augment interleaved pairs using retrieval results from audio-to-audio and audio-to-text methods. In this phase, we freeze the connector and train only the LoRA parameters.

We apply average pooling with a window size of 8 to reduce the number of encoded audio tokens to $N=128$. By maintaining a small $N$, we ensure efficient computation of multiple interleaved audio-text pairs. 
We train the model using the cross-entropy loss computed over all \(K\) interleaved audio–text pairs. Let
\(Z_{K} = \bigl(z_{1}, \dots, z_{K}\bigr)\), \(z_{j} = \bigl(a_{j},\,t_{j}\bigr)\) denote the sequence of \(K\) audio–text pairs, where \(a_{j}\) is the j-th audio and \(t_{j}\) is its corresponding caption.
For the \(k\)-th pair, let \(t_{k,i}\) denote the \(i\)-th token of caption, and let \(t_{k,<i}\) be caption preceding token \(i\). If \(N_k\) is the length of the \(k\)-th caption, the loss over all captions is
\begin{equation}
\mathcal{L}_{\mathrm{CE}}
= -\sum_{k=1}^{K}\sum_{i=1}^{N_k}
  \log p\bigl(t_{k,i}\,\bigm|\,Z_{k-1},\,a_{k},\,t_{k,<i}\bigr).
\label{eq:loss_function}
\end{equation}



\subsection{Generation-assisted multimodal querying (GAMQ)}
In audio captioning, the query is typically provided as a unimodal audio input, whereas the knowledge base consists of multimodal audio-text pairs. To fully leverage the multimodal nature of the knowledge base, we first generate a corresponding text representation for the query audio and then perform multimodal querying using the resulting audio-text pair.

\subsubsection{Audio query to multimodal query transformation:}
To construct a multimodal query, we utilize our base model, trained with progressive learning, to generate a caption \( T_q \) from the given audio query \( A_q \). The caption is produced by simply feeding the query audio into the base model, which generates a textual description that captures the audio's context. This generated caption \( T_q \) serves as the textual component of the multimodal query. By combining the original audio query \( A_q \) with the generated caption \( T_q \), we can perform multimodal querying to retrieve relevant audio-text pairs from the knowledge base.

\subsubsection{Multimodal similarity:}
To choose the top-$K$ candidates from the knowledge base, we need to define a pair-to-pair similarity measure that enables comparison between audio-text pairs.

Let \( \text{Enc}_A (\cdot) \) denote the audio encoder and \( \text{Enc}_T (\cdot) \) denote the text encoder. Given a multimodal query \( (A_q, T_q) \) and a Knowledge Base~(KB) where the $k$'th element is \( (A_k, T_k) \), we utilize KB's pre-computed embeddings for similarity calculation. The similarity $S$ is computed based on the encoded embeddings, as shown below.
\vspace{-2pt}
\begin{align}
\scriptsize
S_A(A_q, A_k) &= \langle \text{Enc}_A(A_q), \text{Enc}_A(A_k) \rangle \label{eq:sa} \\
S_T(T_q, T_k) &= \langle \text{Enc}_T(T_q), \text{Enc}_T(T_k) \rangle \label{eq:st} \\
S &= \alpha \cdot S_A + (1 - \alpha) \cdot S_T \label{eq:s}
\end{align}
%
%
Essentially, we compute the cosine similarity of normalized embeddings between the audio query and the retrieved audio using Eq.~(\ref{eq:sa}), and between the generated text query and the retrieved caption using Eq.~(\ref{eq:st}). The final multimodal similarity score is then determined by Eq.~(\ref{eq:s}), where \( \alpha \) acts as a weighting factor to balance the contributions of audio and text similarities. In our work, we set the weight factor \( \alpha = 0.5 \) to ensure that both modalities are considered equally. To calculate audio and text similarity, we use the pre-trained audio encoder and text encoder of Laion-CLAP.

%
%
%
%
\section{Experiments}
\label{sec:experiment}
We provide the details of the experimental setup in Section~\ref{subsec:exp_setup} and present the three main experimental results in Section~\ref{subsec:rag_performance}.

\subsection{Experimental setup}
\label{subsec:exp_setup}

\textbf{Datasets:}
Three audio captioning datasets, WavCaps~\cite{Mei2023WavCapsAC}, AudioCaps~\cite{kim2019audiocaps}, and Clotho~\cite{drossos2020clotho} version 2.1, are used for training. These datasets are well-known for covering a diverse range of environmental sounds. 
WavCaps is exclusively used in pre-training. We filtered out audios longer than 40 seconds and used 290K audio-text pairs in total. Since WavCaps partially overlaps with AudioCaps and Clotho, we removed audio samples that overlap with the test splits of AudioCaps and Clotho.
The training splits of AudioCaps and Clotho are used for fine-tuning. For the AudioCaps test split, we evaluate 957 audio files out of the original 975 test files due to missing audio links. We balance training by blending AudioCaps and Clotho with an equal number of samples.

\noindent{}\textbf{Knowledge base:}
We combine the training splits of AudioCaps, Clotho, WavCaps, and Auto-ACD to construct our KB. We utilize Laion-CLAP~\cite{2023laionclap} audio encoder and text encoder to pre-compute embeddings for the audio-text pairs in the KB. To accelerate dense retrieval, we utilize the FAISS~\cite{johnson2019billion} library. 

\noindent{}\textbf{Evaluation metrics:}  
To evaluate the performance of our audio captioning model, we employ three established metrics commonly used in audio captioning: CIDEr~\cite{vedantam2015cider}, SPICE~\cite{anderson2016spice}, and SPIDEr~\cite{liu2017improved}. 

\noindent{}\textbf{Training details:}
In stage 1, we train connector of our model using the WavCaps for 5 epochs. We use the AdamW optimizer with a learning rate of \( 1\mathrm{e}{-4} \) and a batch size of 16. A cosine learning rate scheduler is applied.
%
In Stage 2, we train our model for 5 epochs per phase on both AudioCaps and Clotho.
In phase 1 of stage 2, we use AdamW optimizer and cosine learning rate scheduler but reduce the learning rate to \( 1\mathrm{e}{-5} \) and use a batch size of 16. We fine-tune the model using only a single audio-caption pair, without interleaved pairs.
%
In phase 2 of stage 2, we increase the number of interleaved audio-text pairs in a linear manner during training, starting from a single pair and reaching a maximum of 5 pairs by the final epoch. 

\noindent{}\textbf{Retrieval details:}
We use a retrieval process with two sequential steps. First, we retrieve 25 candidate pairs based on audio-to-pair similarity~\cite{yasunaga2022retrieval}. Then, we rank the pairs using a multimodal similarity score, which incorporates both audio and text modalities. Specifically, we use Eq.~(\ref{eq:s}) to compute the final similarity scores and extract the top-$K$ pairs. 


\subsection{Experimental results} 
\label{subsec:rag_performance}
We evaluate our model using AudioCaps and Clotho as primary benchmarks, following common practice in audio captioning. Additionally, we provide results on Auto-ACD~\cite{Sun2023ALD}.
%
%

\noindent{}\textbf{AudioCaps:}
The evaluation results for AudioCaps are presented in Table~\ref{tab:quantitative_evaluation_audiocaps}. For this widely used benchmark, our MQ-Cap outperforms the comparison approaches, including the RAG-based methods RECAP and Audio-Flamingo. In terms of the SPIDEr score, MQ-Cap is the only method that surpasses 0.500, with AutoCap being the next best-performing model, achieving a score of 0.497. We provide four analysis results, including an ablation study, in Section~\ref{sec:analysis}.

\begin{table}[t]
\centering
\captionsetup{skip=5pt}
\caption{Evaluation results for AudioCaps. \( ^* \) denotes the reproduced result for QWen-Audio. \( ^\ddagger \) indicates that Audio-Flamingo was fine-tuned on Clotho.
}
\resizebox{\columnwidth}{!}{%
\begin{tabular}{lcccccc}
\toprule
\textbf{Method}  & \makecell{\textbf{Trainable}\\\textbf{Params}} & \textbf{CIDEr} & \textbf{SPICE} & \textbf{SPIDEr} \\
\midrule
AAC-prefix~\cite{kim2023prefix}                          & 136M  & 0.710 & 0.167 & 0.438  \\
Pengi~\cite{deshmukh2023pengi}                           & 81M   & 0.752 & 0.182 & 0.467  \\
SALMONN~\cite{tang2024salmonn}                           & 33M   & -     & -     & 0.403  \\
QWen-Audio$^*$~\cite{Chu2023QwenAudioAU}                 & 8.34B & 0.640 & 0.174 & 0.407  \\
AL-mixgen~\cite{kim2023exploring}       & 116M  & 0.769 & 0.181 & 0.475  \\
WavCaps~\cite{Mei2023WavCapsAC}         & 170M  & 0.787 & 0.182 & 0.485  \\
EnCLAP~\cite{kim2024ENCLAP}             & 441M  & 0.803 & 0.188 & 0.495  \\
AutoCap~\cite{HajiAli2024TamingDA}      & 1.25B & 0.804 & 0.190 & 0.497  \\
\midrule
\textbf{RAG-based methods:} & & & & & \\
RECAP~\cite{Ghosh2023RECAPRA}                          & 7M   & 0.764 & 0.187  & 0.469 \\
Audio-Flamingo$^\ddagger$~\cite{kong2024audioflamingo} & 2.98B & 0.546  & -     & -      \\
DRCAP~\cite{li2025drcap}                               & 16M & 0.705  & 0.180 & 0.442 \\
MQ-Cap (Ours)                                          & 28M & \textbf{0.845}  &  \textbf{0.194} & \textbf{0.519} \\
\bottomrule \\
\end{tabular}%
}
\label{tab:quantitative_evaluation_audiocaps}
\end{table}

\noindent{}\textbf{Clotho:}
Table~\ref{tab:quantitative_evaluation_clotho} presents the evaluation results for Clotho. MQ-Cap also achieves outstanding performance on this benchmark. 
%
For Audio-Flamingo, it is noted that it was fine-tuned on Clotho and its RAG-based performance was available for AudioCaps, whereas only the no-RAG performance was available for Clotho.
%

\setlength{\textfloatsep}{5pt} 
\setlength{\floatsep}{5pt}      
\setlength{\intextsep}{5pt}     

\begin{table}[t]
\renewcommand{\arraystretch}{1.2} 
\setlength{\tabcolsep}{8pt} %
\captionsetup{skip=5pt}
\caption{Evaluation results for Clotho. $^*$ denotes the reproduced result for SALMONN. \( ^\ddagger \) indicates that Audio-Flamingo was fine-tuned on Clotho. }
\resizebox{\columnwidth}{!}{%
\begin{tabular}{lccccc}
\toprule
\textbf{Method} & \makecell{\textbf{Trainable}\\\textbf{Params}} & \textbf{CIDEr} & \textbf{SPICE} & \textbf{SPIDEr} \\
\midrule
AAC-prefix~\cite{kim2023prefix}                      & 136M  & 0.319 & 0.111 & 0.215  \\
Pengi~\cite{deshmukh2023pengi}                       & 81M   & 0.416 & 0.126 & 0.271  \\
SALMONN$^*$~\cite{tang2024salmonn}                   & 33M   & 0.408 & 0.132 & 0.271  \\
QWen-Audio~\cite{Chu2023QwenAudioAU}                 & 8.34B & 0.441 & 0.136 & 0.288  \\
WavCaps~\cite{Mei2023WavCapsAC}                      & 170M  & 0.462 & 0.133 & 0.297  \\
EnCLAP~\cite{kim2024ENCLAP}                          & 441M  & 0.464 & 0.133 & 0.299  \\
Audio-Flamingo$^\ddagger$~\cite{kong2024audioflamingo}          & 2.98B & 0.465 & -     & -     \\
\midrule
\textbf{RAG-based methods:} & & & & & \\
RECAP~\cite{Ghosh2023RECAPRA}                         & 7M  & 0.323 & 0.116 & 0.221  \\
DRCAP~\cite{li2025drcap}                              & 16M  & 0.438 & 0.133 & 0.285  \\
MQ-Cap (Ours)                                         & 28M  & \textbf{0.496} & \textbf{0.143} & \textbf{0.319} \\
\bottomrule \\
\end{tabular}
}
\label{tab:quantitative_evaluation_clotho}
\end{table}

\noindent{}\textbf{Auto-ACD:}
The Auto-ACD dataset has rarely been used for evaluation in previous works, but its large size makes it an attractive benchmark. Since prior evaluation results for Auto-ACD are not available, we have selected two audio foundation models, QWen-Audio and SALMONN, and evaluated their performance on this dataset. For a fair comparison, we assess performance both with and without applying RAG. The results are presented in Table~\ref{tab:ood_evaluation_autoacd}, where MQ-Cap achieves the best overall scores and the largest incremental gains with RAG, demonstrating strong few-shot captioning performance.

\begin{table}[t]
\captionsetup{skip=5pt}
\caption{Evaluation results for Auto-ACD. Two foundation models are compared. When RAG is used, we apply our generation-assisted multimodal querying. }
\centering
\renewcommand{\arraystretch}{1.2} 
\setlength{\tabcolsep}{8pt} 
\resizebox{\columnwidth}{!}{ 
\begin{tabular}{l l l l l}
\toprule
\textbf{Method}  & \textbf{RAG} & \textbf{CIDEr}                        & \textbf{SPICE} & \textbf{SPIDEr} \\
\midrule
QWen-Audio & No         & 0.436          & 0.150          & 0.293  \\
           & Yes        & 0.591 (+0.155) & 0.181 (+0.031) & 0.386 (+0.093)  \\
\midrule
SALMONN  & No           & 0.415          & 0.156          & 0.286  \\
         & Yes          & 0.588 (+0.173) & 0.179 (+0.023) & 0.383 (+0.097)  \\
\midrule
MQ-Cap (Ours)  & No     & 0.453          & 0.162 & 0.308  \\
               & Yes    & \textbf{0.704 (+0.251)} & \textbf{0.210 (+0.048)} & \textbf{0.457 (+0.149)}  \\
\bottomrule 
\end{tabular}
}
\label{tab:ood_evaluation_autoacd}
\end{table}

\section{Analysis}
\label{sec:analysis}
To better understand the proposed MQ-Cap method, we provide four analysis results in this section. For simplicity, we report only the SPIDEr results on test splits of AudioCaps and Clotho.

\subsection{Ablation study}
\label{subsec:ablation_study}
An ablation study was conducted on the AudioCaps and Clotho to investigate the performance contributions of Progressive Learning~(PL) and Generation-Assisted Multimodal Query~(GAMQ). 
The results are presented in Table~\ref{tab:ablation_performance}. It can be observed that both PL and GAMQ contribute to the overall performance improvement.

\begin{table}[t]
\centering
\captionsetup{skip=5pt}
\caption{Ablation performance in AudioCaps and Clotho. Without PL, the phase 2 was skipped in stage 2. Without GAMQ, conventional audio-to-audio querying was used. } 
\resizebox{0.9\columnwidth}{!}{%
\begin{tabular}{lll}
\toprule
\multirow{2}{*}{\textbf{Method}} & \multicolumn{2}{c}{\textbf{Dataset}} \\ 
\cmidrule(lr){2-3} 
                             & \textbf{AudioCaps} & \textbf{Clotho} \\ 
\midrule
Vanilla model                               & 0.497 & 0.299 \\
Vanilla model + PL                          & 0.507 & 0.312 \\
Vanilla model + PL + GAMQ                   & \textbf{0.519} & \textbf{0.319} \\
%
%
\bottomrule
\end{tabular}
}
\label{tab:ablation_performance}
\end{table}

\subsection{Analysis of four types of queries}
Conventional querying methods in retrieval augmented audio captioning focus on audio-to-audio and audio-to-text retrieval. However, inspired by RA-CM3~\cite{yasunaga2022retrieval} in image captioning, we also evaluate audio-to-pair which considers both modalities.
In Table~\ref{tab:query_design}, we compare four different query choices. For audio-to-pair, only audio is used in the query, but both audio-to-audio and audio-to-text similarities are considered by adopting \( \langle \text{Enc}_A(A_q), \text{Enc}_A(A_k) \rangle + \langle \text{Enc}_A(A_q), \text{Enc}_T(T_k) \rangle \) as the similarity score. Pair-to-pair corresponds to our GAMQ approach, and the similarity score is computed using Eq.~(\ref{eq:s}) with \( \alpha = 0.5 \). 
From the results in Table~\ref{tab:query_design}, we observe that GAMQ achieves the highest SPIDEr scores across all datasets. 
Although GAMQ yields marginal gains on AudioCaps, it delivers significant improvements on Clotho and unseen Auto-ACD, underscoring its robustness, whereas the others fail to perform consistently across benchmarks.

\begin{table}[t]
\centering
\renewcommand{\arraystretch}{1.2} 
\setlength{\tabcolsep}{8pt} 
\captionsetup{skip=5pt} 
\caption{
Analysis of the four types of queries. GAMQ achieves the best SPIDEr performance across both datasets.  
}
\resizebox{\columnwidth}{!}{%
\scriptsize
\begin{tabular}{l c c c}
\toprule
\textbf{Query} & \textbf{AudioCaps} & \textbf{Clotho} & \textbf{Auto-ACD} \\ 
\midrule
Audio-to-Audio    & 0.507 & 0.312 & 0.408 \\
Audio-to-Text     & 0.516 & 0.307 & 0.372 \\
Audio-to-Pair     & 0.517 & 0.308 & 0.416 \\
Pair-to-Pair (GAMQ)  & \textbf{0.519} & \textbf{0.319} & \textbf{0.427} \\
\bottomrule
\end{tabular}
}
\label{tab:query_design}
\end{table}



\subsection{Relevance analysis of the top-5}
\label{subsec:retrieval_similarity} 
\begin{figure}[t]
    \centering
    \includegraphics[width=\columnwidth]{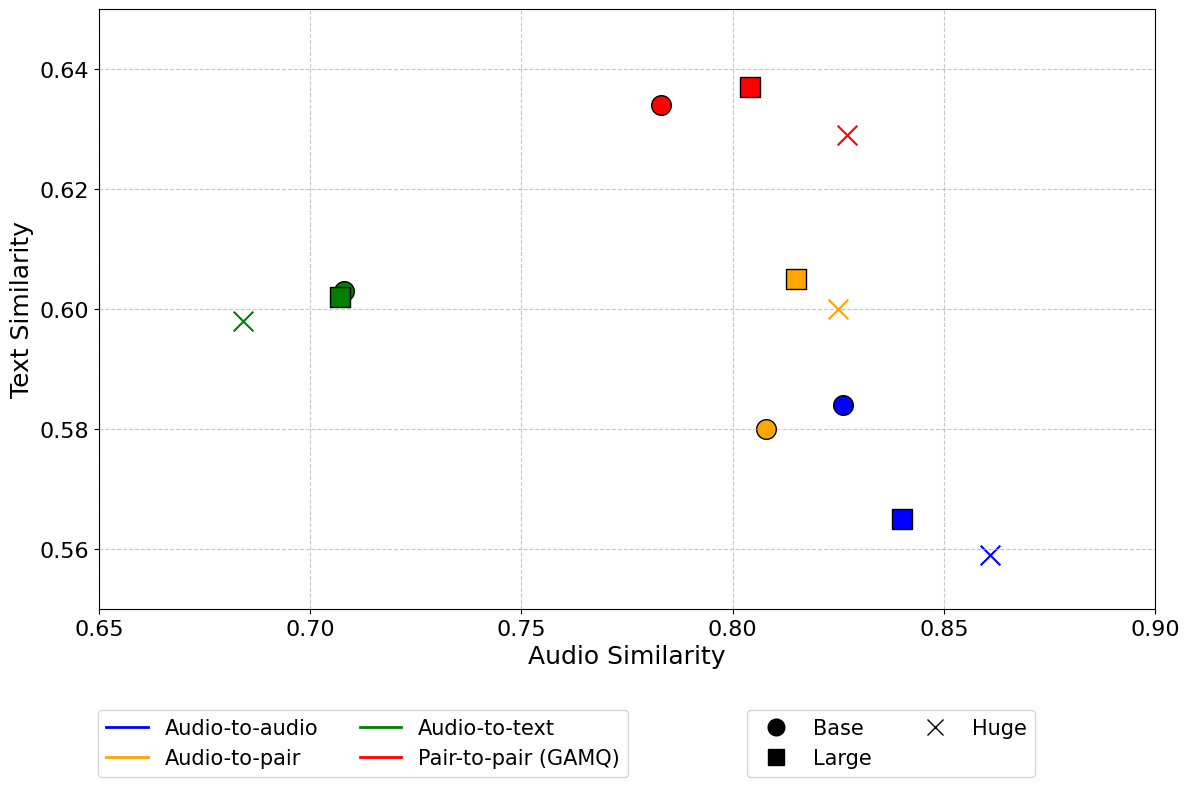}
    \caption{Audio similarity between query audio and the retrieved audios vs. text similarity between ground-truth caption and the retrieved captions on the AudioCaps test split. The knowledge base configurations are as follows: the base KB contains only AudioCaps; the large KB includes AudioCaps, Clotho, and WavCaps; and the huge KB further incorporates Auto-ACD in addition to AudioCaps, Clotho, and WavCaps. }
    \label{fig:similarity_scores}
\end{figure}

To better understand query and knowledge‐base design, we analyzed each audio query and its top-5 retrieved pairs. We computed audio similarity as the average cosine similarity between the query audio and the five retrieved audios, and text similarity as the average cosine similarity between the ground-truth caption and the five retrieved captions.
The results are presented in Figure~\ref{fig:similarity_scores}. For audio similarity, audio-to-audio and audio-to-pair queries achieve slightly better performance than GAMQ. However, GAMQ provides a significant improvement in text similarity. Overall, GAMQ provides a Pareto-optimal balance, achieving the highest text similarity while maintaining competitive audio similarity.



\subsection{Applying multimodal querying with generation assistance to other audio tasks}
\label{subsec:retrieval_eval}
To demonstrate broader applicability of GAMQ, we apply our multimodal querying method to two additional tasks and show how it can improve state-of-the-art performance. The first task, \textit{cross-modal retrieval}\cite{Oncescu2021AudioRW}, aims to identify the best-matching text captions for a given audio query. We enhance retrieval by integrating a text-to-text similarity score with the original audio-text similarity score. As shown in Table~\ref{tab:audio_text_retrieval}, our method, built on state-of-the-art baselines, improves performance by up to 3.1\%.

\begin{table}[b]
\captionsetup{skip=5pt}
\caption{Cross-modal retrieval performance on AudioCaps and Clotho test splits. We evaluate recall for retrieval accuracy. }
\resizebox{\columnwidth}{!}{%

\begin{tabular}{lllll}
\toprule
                        & \multicolumn{2}{l}{\textbf{AudioCaps}} & \multicolumn{2}{l}{\textbf{Clotho}}                   \\
\multirow{-2}{*}{\textbf{Model}} & \textbf{R@1} & \textbf{R@5} & \textbf{R@1} & \textbf{R@5} \\
\midrule
Laion-CLAP~\cite{2023laionclap}            & 44.0 & 79.5 & 21.7 & 45.3 \\
Laion-CLAP~(w/ multimodal query)           & \textbf{45.3} & \textbf{80.3}  & \textbf{23.6} & \textbf{46.6} \\
\midrule
OmniBind~\cite{wang2024omnibind}           & 56.0 & 80.8 & 30.0 & 52.7 \\
OmniBind~(w/ multimodal query)             & \textbf{59.1} & \textbf{83.8}  & \textbf{31.0} & \textbf{54.3} \\
\bottomrule
\end{tabular}
}
\label{tab:audio_text_retrieval}
\end{table}

\setlength{\textfloatsep}{5pt}
\begin{table}[!htbp]
\centering
\captionsetup{skip=5pt}
\caption{Zero-shot classification accuracy across four datasets. \( ^\dagger \)~The baseline performance of Laion-CLAP is reproduced. }
\begin{scriptsize}
\resizebox{\columnwidth}{!}{%
\begin{tabular}{lll}
\toprule
\textbf{Dataset}              & \makecell{\textbf{Laion-CLAP$^\dagger$}} & \makecell{\textbf{Laion-CLAP}\\\textbf{(w/ multimodal query)}} \\
\midrule
ESC-50       & 92.75        & \textbf{95.25 (+2.50)} \\
Urbansound8K & 74.87        & \textbf{78.39 (+3.52)} \\
\midrule
TUT17        & 37.35        & \textbf{38.70 (+1.35)} \\
\midrule
GTZAN        & 66.26        & \textbf{68.07 (+1.81)} \\
\bottomrule
\end{tabular}
}
\end{scriptsize}
\label{tab:zero_shot_pair_to_pair}
\end{table}

The second task, \textit{zero‑shot classification}\cite{2023laionclap}, assigns labels to unseen audio by matching them to predefined class descriptions via cross‑modal similarity. We enhance this process by incorporating a text similarity score using generated captions. As Table~\ref{tab:zero_shot_pair_to_pair} shows, our multimodal querying yields up to a 3.52\% accuracy improvement across sound events, acoustic scenes, and music genres.

\subsection{Analysis of model and retrieval latency}
We measure average captioning latency on an NVIDIA A6000 GPU. Without retrieval, our model processes each audio clip in 0.817 seconds, closely matching Qwen-Audio’s 0.819 seconds. Introducing two interleaved audio–text pairs increases latency to 0.90 seconds for our model and 2.18 seconds for Qwen-Audio, while four pairs incur 0.99 seconds and 3.61 seconds, respectively. Our GAMQ retrieval incurs a generation overhead of 1.07 seconds compared to 0.25 seconds for unimodal audio retrieval, bringing the total captioning latency to 2.06 seconds.
\section{Conclusion}

We presented a retrieval-augmented audio captioning approach that leverages GAMQ and Progressive Learning. Furthermore, we demonstrated that GAMQ is effective for other audio tasks, such as cross-modal retrieval and zero-shot classification.
\label{sec:conclusion}


\section{ACKNOWLEDGEMENTS}
This work was supported by Institute of Information \& communications Technology Planning \& Evaluation (IITP) grant funded by the Korea government (MSIT) [NO.RS-2021-II211343, Artificial Intelligence Graduate School Program (Seoul National University)], [No. RS-2023-00235293, Development of autonomous driving big data processing, management, search, and sharing interface technology to provide autonomous driving data according to the purpose of usage]).

\bibliographystyle{IEEEtran}
\bibliography{mybib}

\end{document}